\documentclass[aps,plr,twocolumn,showpacs,superscriptaddress]{revtex4-2}
\usepackage{amsmath}
\usepackage{amssymb}
\usepackage{graphicx}
\usepackage{subfigure}
\usepackage{color}
\usepackage{float}
\usepackage{graphics}
\usepackage{hyperref}
\usepackage{txfonts}
\usepackage{tabularx}
\usepackage{array}
\usepackage{booktabs}

\begin{document}
\title{Superexchanges and Charge Transfer in the La$_3$Ni$_2$O$_7$ Thin Films}

\author{Yuxun Zhong}
\affiliation{State Key Laboratory of Optoelectronic Materials and Technologies, Guangdong Provincial Key Laboratory of Magnetoelectric Physics and Devices, Center for Neutron Science and Technology, School of Physics, Sun Yat-Sen University, Guangzhou 510275, China}

\author{Wéi Wú}
\email{wuwei69@mail.sysu.edu.cn}
\affiliation{State Key Laboratory of Optoelectronic Materials and Technologies, Guangdong Provincial Key Laboratory of Magnetoelectric Physics and Devices, Center for Neutron Science and Technology, School of Physics, Sun Yat-Sen University, Guangzhou 510275, China}

\author{Dao-Xin Yao}
\email{yaodaox@mail.sysu.edu.cn}
\affiliation{State Key Laboratory of Optoelectronic Materials and Technologies, Guangdong Provincial Key Laboratory of Magnetoelectric Physics and Devices, Center for Neutron Science and Technology, School of Physics, Sun Yat-Sen University, Guangzhou 510275, China}

\begin{abstract}
The recent discovery of  ambient-pressure superconductivity with $T_c$ above 40 K in La$_3$Ni$_2$O$_7$ thin films represents a significant advance in the field of nickelate superconductor. Motivated by the experimental reports, here we study an 11-band $d-p$ Hubbard model with tight-binding parameters derived from \textit{ab initio} calculations, using large scale determinant quantum Monte Carlo and cellular dynamical mean-field theory. Our results reveal that the major superexchange couplings in La$_3$Ni$_2$O$_7$ thin films can be substantially weaker than in the bulk material at 29.5 Gpa. Specifically, the out-of-plane antiferromagnetic correlation between Ni$-d_{3z^2-r^2}$ orbitals is reduced by about 27\% in film, while   the in-plane magnetic correlations remain largely unaffected. We evaluate the corresponding antiferromagnetic coupling constants,  $J_{\perp}$ and $J_{\parallel}$ using perturbation theory. With regard to charge transfer properties, we find that the biaxial compression in films reduces charge transfer gap. We also resolve the orbital distribution of doped holes and electrons among the in-plane (Ni$-d_{x^2-y^2}$ and O$-p_x/p_y$) and the out-of-plane (Ni$-d_{3z^2-r^2}$ and O$-p_z$) orbitals, uncovering a pronounced particle-hole asymmetry. Theses findings lay a groundwork for the study of low-energy  $t-J$ model of  La$_3$Ni$_2$O$_7$ films and provide key insights into the understanding of physical distinctions between the film and bulk bilayer nickelates.
\end{abstract}
\maketitle

\textit{Introduction.-} The discovery of pressure-induced superconductivity in Ruddlesden-Popper (RP) nickelates, first in bilayer La$_3$Ni$_2$O$_7$ ($T_c\approx$ 80 K) \cite{Sunsignatures2023} and subsequently in trilayer La$_4$Ni$_3$O$_{10}$ ($T_c=20\sim30$ K) \cite{zhu_superconductivity_2024,cpl_4310}, has stimulated extensive research. After numerous theoretical investigations 
\cite{luo2023bilayer,zhang2023electronic,lechermann2023electronic,shilenko2023correlated,chen2025charge,ouyang2024absence,liu2023s,heier2024competing,zhang2024structural,zhang2024electronic,christiansson2023correlated,wu2024superexchange,tian2024correlation,ryee2024quenched,shen2023effective,lu2024interlayer,oh2023type,qu2024bilayer,yang2024strong,liao2023electron,fan2024superconductivity,kaneko2024pair,luo2024high,yang2023possible,sakakibara2024possible,cao2024flat,yang2023interlayer,huang2023impurity,zhang2023trends,jiang2024pressure,zhang2024strong,cpl_41_7_077402} and experimental studies 
\cite{cpl_4310,yang_orbital-dependent_2024,zhang2024high,hou2023emergence,PhysRevB.110.134520,PhysRevX.14.011040,liu2024electronic,Li_2025}, various protocols have been proposed to obtain superconductivity without external pressure, including both chemical substitution
\cite{Pan__2024,PhysRevB.108.165141,li2025ambientpressuregrowthbilayer,Zhu_2025}and strain engineering
\cite{osada2025strain,wang2025electronicstructurecompressivelystrained,bhatt2025resolvingstructuraloriginssuperconductivity,Liu_2025}.
Recent experiments have achieved this goal in thin films of La$_3$Ni$_2$O$_7$ (LNO) \cite{La3Ni2O7film} and (La,Pr)$_3$Ni$_2$O$_7$ \cite{(LaPr)3Ni2O7film}, reporting superconductivity at ambient pressure with $T_c$ exceeding 40K.  This advance overcomes the experimental difficulties associated with high-pressure measurements, facilitating the use of sophisticated probes like the angle-resolved photoemission spectroscopy (ARPES) or scanning tunneling microscopy (STM). 

Superconductivity in these experiments is achieved by growing LNO films on substrates with reduced lattice constants (such as LaAlO$_3$ (LAO) or SrLaAlO$_4$ (SLAO)), which imposes a strong in-plane biaxial compressive strain~\cite{shao2025bandstructurepairingnature}. This strain can suppress the octahedral tilting in the \textit{Amam} phase of bulk materials at ambient pressure~
\cite{cao2025strainengineeredelectronicstructuresuperconductivity,osada2025strain,samanta2025inplanenionibondangles},  stabilizing a tetragonal structure in  films analogous to the high-pressure \textit{Fmmm} or \textit{I4/mmm} phase of superconducting bulk materials.
Given this structural similarity, the low-energy physics is often assumed to be comparable between high-pressure bulk materials and ambient-pressure films. Yet, whether their effective Hamiltonians are truly equivalent is untested. A central unresolved issue, therefore, is to what extent the modified lattice parameters in the films alter the strength of magnetic superexchange couplings and the orbital distribution of the charge carriers. These two factors are believed to govern the system’s superconducting properties.
\par
To address above question, we study an 11-band $d-p$ Hubbard model that includes four $3d_{x^2-y^2}$ / $3d_{3z^2-r^2}$ orbitals of nickel and seven most relevant $2p$ orbitals of oxygen in the NiO$_2$ bilayer per unit cell \cite{hu2025electronic}. We perform numerically exact determinant quantum Monte Carlo (DQMC) simulations \cite{PhysRevD.24.2278,assaad_world-line_2008} and cellular dynamical mean-field theory (CDMFT) calculations \cite{RevModPhys.68.13,RevModPhys.77.1027} in the normal state of the system. 
Our results suggest that LNO thin film system exhibits weaker inter-layer (IT) antiferromagnetic (AFM) couplings between $d_{3z^2-r^2}$ orbitals, while the strength of the intra-layer (IR) AFM couplings between $d_{x^2-y^2}$ orbitals remain nearly unchanged comparing to the bulk LNO. Furthermore, we analyze the charge-transfer properties \cite{PhysRevX.10.021061} in the Zaanen-Sawatzky-Allen (ZSA) scheme \cite{1985chargetransfer}, which has been shown crucial for understanding $T_c$ in cuprate superconductors \cite{CuprateRUAN20161826,CuprateChargeTransfer,Weber_2012cuprate,Shane2022pairingmechanism_of_copper-oxide,rybicki_perspective_2016}. We reveal that the holes introduced by doping the pristine system with heterovalent elements are distributed approximately evenly between the in-plane and out-of-plane orbitals, whereas for electron doping, it favors a roughly 3:1 occupancy ratio. By comparing with the experimental doping phase diagram \cite{hao2025superconductivityphasediagramsrdoped} of superconducting La$_{3-x}$Sr$_x$Ni$_2$O$_7$ films, we argue that the occupation number of $d_{3z^2-r^2}$ is crucial for determining the type of superconductivity \cite{mo2025intertwinedelectronpairingbilayer}, namely, whether it is driven by Hund's coupling $J_H$ \cite{qu2024bilayer,oh2023type,lu2024interlayer,PhysRevB.111.144514} or governed by hybridization $V$ between $e_g$ orbitals \cite{PhysRevB.111.035108}.

\begin{figure}[t]
    \centering
    \includegraphics[width=3.4 in]{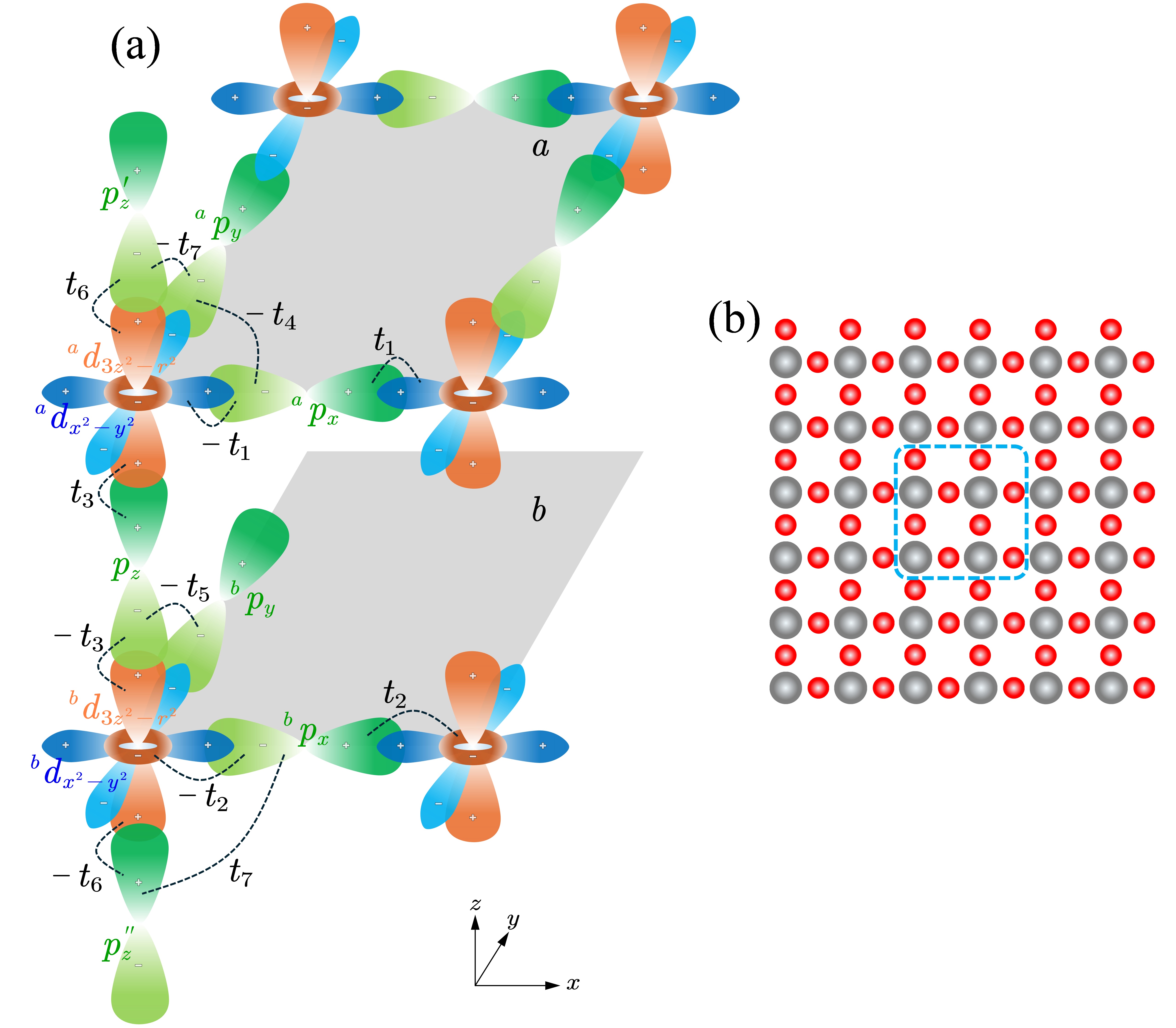}
    \caption{Schematic of the 11-band bilayer LNO thin film lattice model with the seven major hopping parameters and the illustration of the two-dimensional NiO$_2$ lattice for DMQC or CDMFT simulations. \textbf{(a)}. The orange, blue and green electron clouds denote Ni-3$d_{3z^2-r^2}$, Ni-3$d_{x^2-y^2}$ and O-2$p$ orbitals. For brevity, some of the orbitals on plane $b$ is not displayed. Four hopping processes between Ni-$d$ and O-$p$ orbitals predominantly contribute to superexchange interactions between Ni-$d$ orbitals are: $t_1=-1.4795,t_2=0.6605,t_3=-1.4565,t_6=1.2045$. The site energies are: $\varepsilon_{d_{x^2-y^2}}=-0.962,\varepsilon_{d_{3z^2-r^2}}=-1.070,\varepsilon_{p_{x}/p_{y}}=-4.650,\varepsilon_{p_{z}}=-4.078,\varepsilon_{p_{z}^{'}}=\varepsilon_{p_{z}^{''}}=-2.977$ \cite{hu2025electronic}. The superexchanges between the intra-layer $d_{x^2-y^2}$ and $d_{3z^2-r^2}$ orbitals (not shown here) vanishes due to symmetry. Here $t_4=0.5675,t_5=0.4315,t_7=0.4115$ denote electrons hopping processes between O-$p$ orbitals.  \textbf{(b)}. Dark balls denote nickel positions while light balls show oxygen positions in the NiO$_2$ plane. The 6 $\times$ 6 unit cells used in DQMC simulations is shown here directly, while the 2 $\times$ 2 CDMFT effective cluster is outlined by a blue dashed rectangle.  For clarity, only one NiO$_2$ layer is shown here.}
    \label{fig1}
\end{figure}

\textit{Model and method.-}
 To investigate the superexchange couplings in LNO films, we consider a high-energy effective model including eleven Ni-$d$ and O-$p$ orbitals. The Halmiltonian is given as:
\begin{equation}
\begin{aligned}
    H = &\sum_{ijab\alpha\beta\sigma}t_{ia\alpha,jb\beta}c_{ia\alpha\sigma}^{\dagger}c_{jb\beta\sigma}+\sum_{a\alpha\sigma}(\varepsilon_{a\alpha}-\mu)n_{a\alpha\sigma}-\sum_{ia\alpha\sigma}E_{\mathrm{DC}}^{a\alpha}n_{ia\alpha\sigma}^{d}\\
    &+U\sum_{ia\alpha}n_{ia\alpha\uparrow}^{d}n_{ia\alpha\downarrow}^{d}+\sum_{iab\alpha\beta\sigma\sigma'}(V-J_H\delta_{\sigma\sigma'})n_{ia\alpha\sigma}^{d}n_{jb\beta\sigma'}^{d},
\end{aligned}
\label{11 orbitals H}
\end{equation} 
where i/j, a/b, $\alpha/\beta$, labels lattice sites, NiO$_2$ layers and orbitals respectively, $\sigma$ denotes spin degree of freedom. The superscript $d$ indicates that the Hubbard-$U$ is applied only to Ni-$3d$ orbitals. Here, $t_{ijab\alpha\beta}$ are electron hopping integrals, $\varepsilon_{a\alpha}$ stands for on-site energy of orbital $\alpha$ in layer $a$. $U$ is the Hubbard repulsion between two electrons on the same $d$ orbital, and $V=U-2J_H$ is for that on two different orbitals \cite{PhysRevLett.86.5345}, where $J_H$ is the Ising-type Hund's coupling \cite{wu2024superexchange}. Here, the pair-hopping and spin-flip terms in Hund's coupling are neglected. The double-counting term $E_{\mathrm{DC}}$ is subtracted in both the DQMC and CDMFT simulations. We adopt Held's formula \cite{held2007electronic} for the double-counting, $E_{DC}=\frac{1}{3}(U+2V-J_H)(n^0_d-0.5)$ where $n_d^0$ is the occupation number of $d-$orbital in the non-interacting limit. In our case $n_d^0\approx2.14$. The hopping parameters and on-site energies are adopted from a DFT calculation on the double-stacked LNO thin film in Ref. \cite{hu2025electronic}. See also \autoref{fig1} (a) and its caption for all the hopping parameters and on-site energies. Following DFT results \cite{hu2025electronic}, we set the chemical potential $\mu = 0$ in the Hamiltonian to represent the pristine LNO thin films at ambient pressure. For the purpose of exploring different doping regimes, we adjust $\mu$ and define the hole concentration 
$n_h=\frac{1}{4}(22-\sum_{a\alpha\sigma}n_{a\alpha\sigma})$, representing the average number of holes per $d$-orbital per site. Unless stated otherwise, we use $U=7$ eV as our default parameters \cite{wu2024superexchange,PhysRevX.10.021061} for Ni $e_g$ orbitals. For the comparative calculations on bulk LNO, we employ a set of parameters for the 11-bands model \cite{luo2023bilayer,wu2024superexchange}.
\par
We work on a 2D square lattice containing up to 6 $\times$ 6 $\times$ 11 = 396 orbitals for DQMC simulation and an effective impurity model with 2 $\times$ 2 $\times$ 11 = 44 orbitals for CDMFT study, as shown in \autoref{fig1} (b). Due to the fermionic sign problem in DQMC, our simulations are restricted to a high-temperature regime, where the lowest temperature $T = 0.25$ (in eV units) is achieved with an average sign $\langle s \rangle \approx 0.79$. The CDMFT is employed as a complementary method accessing low-temperature regime, where $T\sim 0.08$ can be reached while maintaining an average sign $\langle s \rangle \approx 0.56$ in the impurity solver.


\textit{Superexchanges.-} 
We first discuss the magnetic exchange couplings in LNO thin films and compare them with bulk LNO. As demonstrated in \autoref{fig1} (a), four hopping processes ($t_1$, $t_2$, $t_3$ and $t_6$) contribute to the main superexchange couplings, with the corresponding coupling constants estimated as $J_\parallel \sim 0.084$ eV and $J_\perp \sim 0.135$ eV  at $U=7 eV$ in the atomic limit, see \autoref{table1}. In the atomic limit, the charge-transfer picture for spin-1/2 electron system suggests that the energy difference between the spin singlet and spin triplet states composed of the two Ni $d$-orbitals can be written as,
\begin{align}
    J &= E_{\uparrow\uparrow}-E_{\uparrow\downarrow} \notag \\
    &= 2|t_{pd_{1}}t_{pd_{2}}|^2\left(\frac{1}{U+\Delta_{pd_{2}}-\Delta_{pd_{1}}}\left(\frac{1}{U+\Delta_{pd_{1}}}\right)^2 \right. \notag \\
    &\quad \left. {}+\frac{1}{U+\Delta_{pd_{1}}-\Delta_{pd_{2}}}\left(\frac{1}{U+\Delta_{pd_{2}}}\right)^2 \right. \notag \\
    &\quad \left. {}+\frac{1}{2U+\Delta_{pd_{1}}+\Delta_{pd_{2}}}\left(\frac{1}{U+\Delta_{pd_{1}}}+\frac{1}{U+\Delta_{pd_{2}}}\right)^2\right),
    \label{eq_J}
\end{align}
where $t_{pd_{1,2}}$ denotes the hopping between the Ni-$d_{1,2}$ and the O-$p$ orbitals. The energy difference between the lower Hubbard band of the $d$-orbitals and the O-$p$ orbital is given by: $ \Delta_{pd_{1,2}} = \varepsilon_{d_{1,2}} - \varepsilon_p$, with the double-counting term included in $\varepsilon_{d_{1,2}}$. \autoref{eq_J} represents the second-order approximation result from perturbation theory, which describes the superexchange between two neighboring Ni-$3d$ orbitals in the atomic limit. Our result of $J$ for various $U$ can be found in \autoref{table1}, where one sees that, as expected, the AFM coupling coefficient $J$ decreases monotonically with increasing $U$. For instance, the interlayer couplings $J_{\perp}$ drops from 0.325 eV for $U=3$ eV to 0.135 eV for $U=7$ eV.
\begin{figure}[!tbp]
    \centering
    \includegraphics[width=3.4 in]{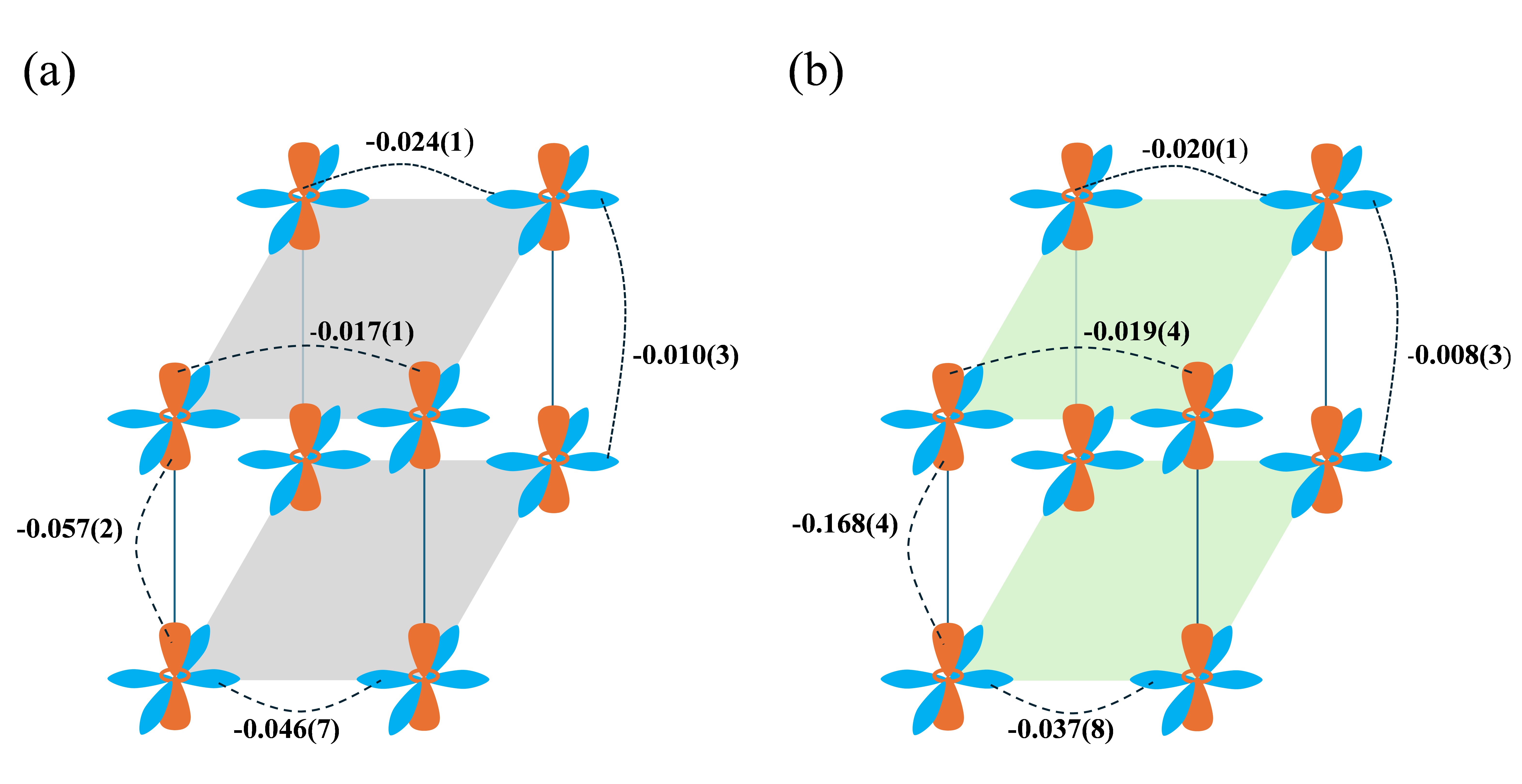}
    \caption{The spin-spin correlation function $\langle S_{ia\alpha}\cdot S_{jb\beta}\rangle$ for four neighboring $d$-orbitals is shown in numbers to demonstrate the relative strength of the antiferromagnetic superexchange couplings in the system. The orange and blue symbols represent $d_{3z^2-r^2}$ and $d_{x^2-y^2}$ orbitals respectively. We  have used the specific values $U = 7$ and $J = 0.15U$, very similar to those obtained from constrained RPA \cite{christiansson2023correlated}.  \textbf{(a)} are from DQMC at half-filling ($n_h=1$, $\mu=1.5$) at $T = 0.25$, and \textbf{(b)} are from CDMFT at $\mu = 0$ ($n_h \approx 1.234$). Note that the magnetic correlations between the on-site inter-layer $d_{x^2-y^2}$ and $d_{3z^2-r^2}$ orbitals, which are due to Hund’s coupling, are not shown here.}
    \label{fig2}
\end{figure}

\begin{table}[b]
  \centering
  \setlength{\tabcolsep}{20pt}
  \caption{The value of superexchanges $J$ at various Hubbard $U$ from \autoref{eq_J}. The vertical inter-layer AFM coupling $J_{\perp}$, intra-layer AFM coupling $J_{\parallel}$ and their ratio $J_\parallel/J_\perp$ are shown. The double counting term used here is the same as in the DQMC and CDMFT simulations: $E_{DC} = 0.473U$ and $E_{DC} = 0.589U$ for $d_{x^2-y^2}$ and $d_{3z^2-r^2}$ orbitals, respectively. The calculations are performed up to sixth-order precision.} 
  \begin{tabular}{@{} c c c c @{}}
    \toprule
    {$U$/eV} & {$J_{\perp}$/eV} & {$J_{\parallel}$/eV} & {$J_\parallel/J_\perp$} \\
    \midrule
    3 & 0.325 & 0.243 & 74.6\% \\
    4 & 0.263 & 0.181 & 68.6\% \\
    5 & 0.207 & 0.137 & 66.1\% \\
    6 & 0.165 & 0.106 & 64.5\% \\
    7 & 0.135 & 0.084 & 62.3\% \\
    \bottomrule
  \end{tabular}
  \label{table1}
\end{table}
\par
In DQMC calculations, the magnetic correlation functions can be exactly measured at high temperatures. \autoref{fig2} (a) and (b) show the magnitudes of the spin correlation function $\langle S_{ia\alpha}\cdot S_{jb\beta}\rangle$ between neighboring $d$- orbitals at $T = 0.25$ and half-filling from DQMC, and $\mu=0$ at $T=0.125$ from CDMFT, respectively. Here $S_{ia\alpha}$ is spin operator with site index $i$, layer $a$ and orbital $\alpha$. This results depicts the relative strengths of the major exchange couplings in the LNO thin film.  
At $T = 0.25$ and half-filling, like in bulk LNO, the dominant AFM exchange coupling in films is also the IT $d_{3z^2-r^2}-d_{3z^2-r^2}$ coupling , with a value of $\langle S\cdot S\rangle = - 0.057(2)$. The next strongest AFM exchange comes from IR nearest-neighboring (NN) $d_{x^2-y^2}-d_{x^2-y^2}$ orbitals, with $\langle S\cdot S\rangle = - 0.046(7)$. Comparing to that in bulk, the in-plane coupling becomes significant in thin films, reaching 81\% of the AFM exchange between IT $d_{3z^2-r^2}-d_{3z^2-r^2}$ orbitals. The same ratio in bulk LNO it is only about 60\% \cite{wu2024superexchange}. On the other hand, the strengths of IR NN $d_{x^2-y^2}-d_{3z^2-r^2}$ and $d_{3z^2-r^2}-d_{3z^2-r^2}$ AFM correlations are $\langle S\cdot S\rangle = - 0.024(1)$ and $\langle S\cdot S\rangle = - 0.017(1)$, respectively. These couplings are relatively small due to a weak  hopping between $d_{3z^2-r^2}$ and $p_x/p_y$ orbitals. Notably, the IT $d_{x^2-y^2} - d_{x^2-y^2}$ exchange coupling within a unit cell is the weakest among these magnetic couplings.  
In \autoref{fig2} (b), CDMFT results at $T = 0.125$ and $\mu = 0$ also suggests that the IT $d_{3z^2-r^2}-d_{3z^2-r^2}$ coupling is predominant in the system, which has a value of $\langle S\cdot S\rangle = - 0.168(4)$. Notably, in CDMFT at $\mu=0$, the strength of the correlation between IT $d_{x^2-y^2}-d_{x^2-y^2}$ orbitals is again the weakest of the five magnetic couplings. In essence, here it is clear that the IT $d_{3z^2-r^2}-d_{3z^2-r^2}$ magnetic correlation dominates at different $\mu$ (or fillings) , suggesting that $d_{3z^2-r^2}$ orbitals should be the key components responsible for the correlation-driven Superconductivity.

\begin{figure}[!tbp]
    \centering
    \includegraphics[width=3.4 in]{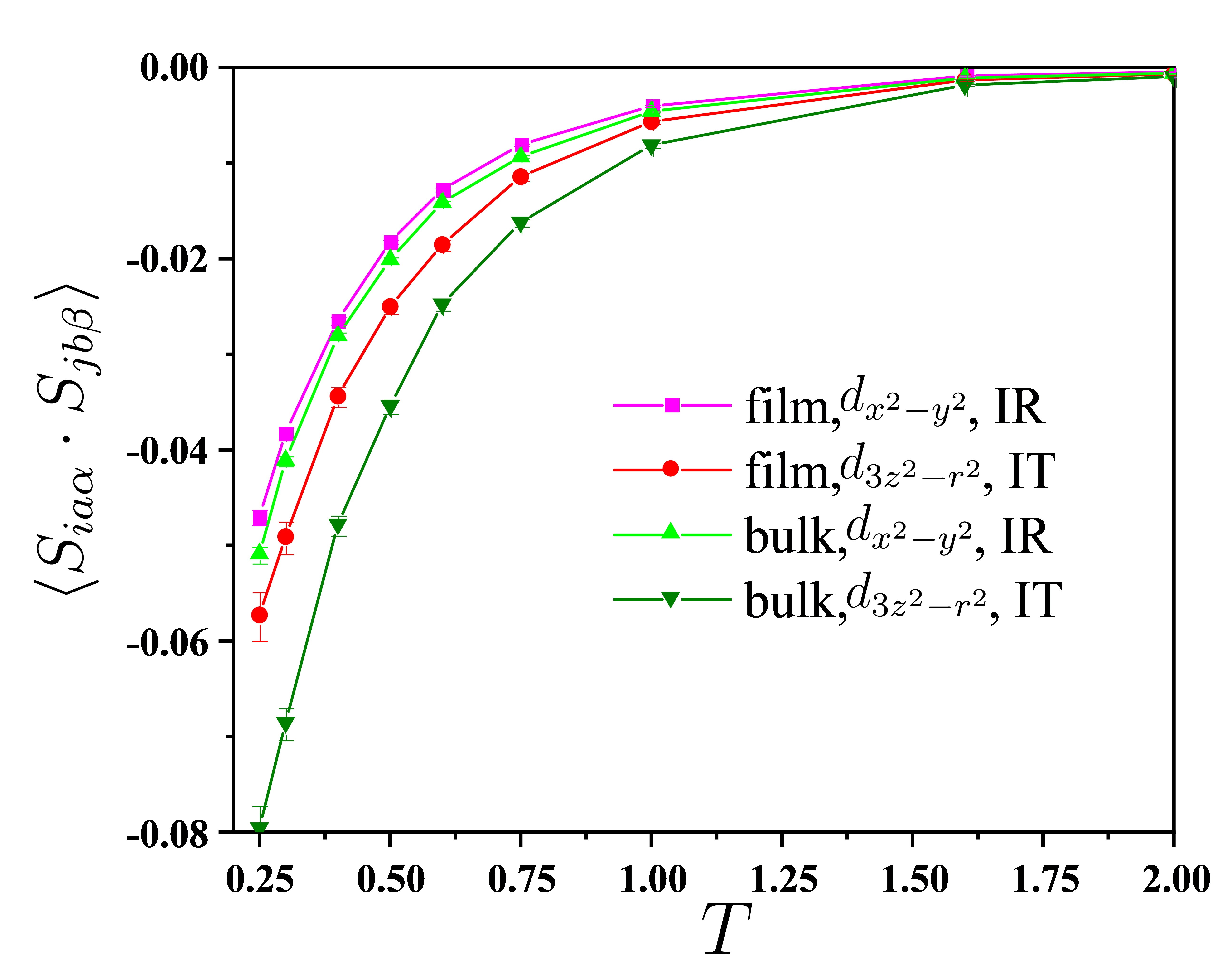}
    \caption{$\langle S_{ia\alpha}\cdot S_{jb\beta}\rangle$ between pairs of nearest-neighboring (NN) d-orbitals as a function of temperature $T$ at pristine configuration, as obtained from DQMC simulations. The squares indicate results for the intra-layer (IR) $d_{x^2-y^2}-d_{x^2-y^2}$ correlations, while the circles show the results for the inter-layer (IT) $d_{3z^2-r^2}-d_{3z^2-r^2}$ correlations.}
    \label{fig3}
\end{figure}
\par
\autoref{fig3} presents the temperature dependence of AFM correlations $\langle S_{ia\alpha}\cdot S_{jb\beta}\rangle$ obtained from DQMC simulations at $\mu=0$. It is clear that bulk LNO system exhibits stronger AFM correlations than those in the thin film for both IR and IT components. This overall suppression of magnetic couplings provides a plausible explanation for the reduced superconducting transition temperature $T_c$ in thin films. A key distinction in the suppression of magnetic couplings should be noticed: the reduction of magnetic correlations for the IT $d_{3z^2-r^2}-d_{3z^2-r^2}$ component is much more prominent than that for the IR NN $d_{x^2-y^2}-d_{x^2-y^2}$ component. The latter appears almost identical between film and bulk systems.  Furthermore, it can be seen that the IT $d_{3z^2-r^2}-d_{3z^2-r^2}$ magnetic correlation in thin film is not significantly stronger than the IR NN $d_{x^2-y^2}-d_{x^2-y^2}$ component. This contrasts sharply with the bulk case, where the IT $d_{3z^2-r^2}-d_{3z^2-r^2}$ correlation is much stronger than its IR $d_{x^2-y^2}-d_{x^2-y^2}$ counterpart\cite{wu2024superexchange}. These results suggests that both the IT $d_{3z^2-r^2}-d_{3z^2-r^2}$   and the IR $d_{x^2-y^2}-d_{x^2-y^2}$ component of AFM correlations may play crucial roles in the ambient-pressure superconductivity observed in LNO thin films. 

\begin{figure}[tbp]
    \centering
    \includegraphics[width = 3.4 in]{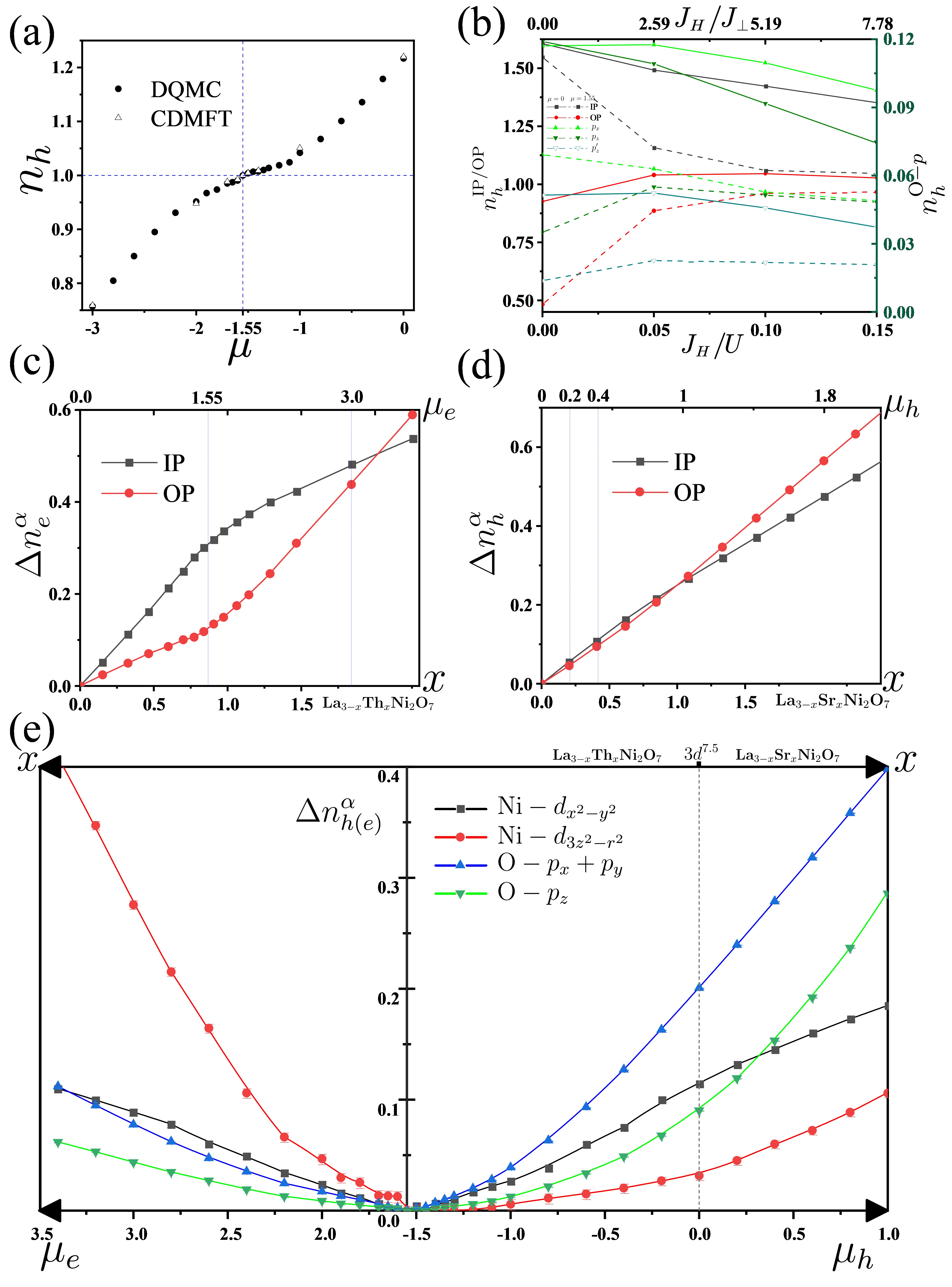}
    \caption{Charge transfer properties and charge carrier evolution of the  La$_3$Ni$_2$O$_7$ thin film. \textbf{(a)}. DQMC and CDMFT result of hole concentration $n_h$ as a function of hole chemical potential $\mu_h$. An inflection point suggests the opening of a charge transfer gap at half-filling when the hole chemical potential $\mu_h$ approaches $\mu_h \sim -1.55$. Here $T=0.3$. \textbf{(b)}. Carrier concentration as a function of Hund's coupling $J_H$ for each orbital in the LNO thin film under half-filling (dashed line) and pristine configuration ($\mu=0$, solid line) conditions by CDMFT at $T=0.125$. The left axis represents the in-plane and out-of-plane hole concentrations, while the right axis shows the hole concentration of oxygen orbitals. \textbf{(c)} and \textbf{(d)} show electron(hole) concentration variation $\Delta n_{e(h)}^\alpha$ as functions of chemical potential $\mu_{e(h)}$ and $\text{Sr}^{2+}(\text{Ce}^{4+})$ nominal doping levels $x$, respectively. Here, $\Delta n_h^{\mathrm{IP}} = \Delta n_h^{d_{x^2-y^2}} + \Delta n_h^{p_x} + \Delta n_h^{p_y}$ counts the number of holes in the in-plane $d_{x^2-y^2}$ orbital and $p_x/p_y$ orbitals combined, while $\Delta n_h^{\mathrm{OP}} = \Delta n_h^{d_{3z^2-r^2}} + 0.5(\Delta n_h^{p_z} + \Delta n_h^{p_z'})$ counts the number of holes in the out-of-plane. Note that the $\Delta n_h^\alpha$ here represents the difference in hole concentration relative to the pristine LNO thin film (Ni-$3d^{7.5}$), i.e. Hole concentration variation $\Delta n_{h}^\alpha \equiv n_{h}(\mu) - n_{h}(\mu_h = 0)$. When Sr doping level $x$ = 0.203, the nominal hole content in La$_{3-x}$Sr$_x$Ni$_2$O$_7$ corresponds to $\mu_h=0.2$, while at $\mu_h$ of 0.4, the equivalent Sr doping level is $x$ = 0.406. At this doping level, superconductivity is no longer observed according to Ref. \cite{hao2025superconductivityphasediagramsrdoped}. \textbf{(e)}. The electron concentration variation $\Delta n_e^\alpha$ as a function of the electron chemical potential $\mu_e$ in the regime where $\mu_e > 1.55$ (corresponding to electron doping with $n_e > 1$) and hole concentration variation $\Delta n_h^\alpha$ in the regime where hole doping with $n_h = 2-n_e > 1$($n_{e/h} = \sum_{a,\alpha}n^{a\alpha}_{e/h}$), respectively. Here, $\alpha$ denotes distinct $\alpha$-orbitals and charge concentration variation $\Delta n_{e/h}^\alpha \equiv n_{e/h}(\mu) - n_{e/h}(\mu_e = 1.55)$. Results in (c), (d) and (e) above are from DQMC simulations at $T = 0.3$.}
    \label{fig4}
\end{figure}
\textit{Charge transfer properties.-} 
 We now examine the charge transfer properties within the 11-band Hubbard model. \autoref{fig4} illustrates the evolution of the hole/electron concentrations across different orbitals as functions of chemical potentials $\mu$ and Hund's coupling $J_H$. In  \autoref{fig4}(a), hole concentration $n_h$ is shown as a function of $\mu$, where an inflection point near $n_h = 1$ ($\mu_h \sim -1.55$) with strongly reduced compressibility ($\partial n_h/\partial\mu_h$) can be seen, suggesting the opening of a charge gap at half-filling ($n_h = 1$). For the temperatures we have accessed ($T>0.125$), the $\mu-n_h$ curve is not completely flat, indicating that the chager gap is not fully open. Notably, the reduced flatness of the $\mu-n_h$ curve around $n_h = 1$ for the film (compared to the bulk at this temperature \cite{wu2024superexchange}) should point to a smaller charge transfer gap in the system.

 \autoref{fig4} (e) presents orbital-resolved distribution of doped charge carriers. As the hole chemical potential $\mu_h \equiv -\mu$ or
electron chemical potential $\mu_e \equiv \mu$ are tuned from half-filling, doped carriers enter into different orbitals. Specifically, we see that doped holes predominantly occupy O-$p$ orbitals ( Blue and Green lines in the right panel of \autoref{fig4} (e)), while doped electrons primarily occupy Ni-$d$ orbitals ( Red and black curves in the left panel of \autoref{fig4} (e)). This observation confirms the charge-transfer nature of the correlated states in the system\cite{1985chargetransfer,CuprateRUAN20161826}. Moreover, we observe that the $d_{x^2-y^2}$ orbital contains a higher concentration of doped holes than the $d_{3z^2-r^2}$ orbital in the case of hole doping ($\mu_h > -1.55$). This suggests that  the $d_{3z^2-r^2}$ orbitals are more correlated comparing to the $d_{x^2-y^2}$ orbital under hole doping.

 To shed light on low-energy effective  models integrating out oxygen degrees of freedom, we group the Ni-$3d$ and O-$2p$ orbitals into two categories: (i) the out-of-plane (OP) correlated orbitals, including $d_{3z^2-r^2}$, $p_z$ and $p_z'$ and (ii) in-plane (IP) correlated orbitals comprising $d_{x^2-y^2}$ and $p_x/p_y$ orbitals. As shown in \autoref{fig4} (d), additional holes doped into  LNO film  distribute nearly equally between IP and OP orbitals. In contrast,  for electron doping around $\mu_e \sim 0$,  the doped carriers favor IP orbitals over OP orbitals by a ratio of approximately 3:1. This stark asymmetry between electron and hole doping  (1:1 vs. 3:1) suggests a dopant-type dependent superconducting phase diagram. This is particularly relevant to experiment, where Sr-doping introduces holes \cite{hao2025superconductivityphasediagramsrdoped}, while Thorium (Th) doping could introduce electrons.
 \par

The data in \autoref{fig4}(d) also show a significant difference in carrier response: under identical chemical potential changes, the hole variation in LNO films is about twofold greater than in the bulk \cite{wu2024superexchange}. This indicates that LNO thin films possess enhanced charge transfer capabilities and, consequently, a smaller charge transfer gap $\Delta$ \cite{zhong2025epitaxialstraintuningelectronic}. Our conclusion aligns with recent spectroscopic experiments \cite{XASfilm}, which propose the existence of two types of Zhang–Rice singlets in LNO films and suggest the in-plane charge transfer properties are analogous to those of cuprate superconductors.

Finally, we find that the presence of Hund's coupling $J_H$ at half-filling strongly in general suppresses orbital polarization, hence tends to stabilize a high-spin $S=1$ state, as illustrated by the dashed lines  in \autoref{fig4} (b), which follows the Hund's first law. However, for the pristine configuration ($\mu=0$), the influence of the $J_H$ coupling is significantly weaken. In this case, holes retain a preferential occupation of the in-plane orbitals, as indicated by the solid lines in \autoref{fig4} (b).

\textit{Discussion and conclusion.-} The observation of superconductivity in La$_3$Ni$_2$O$_7$ thin films at ambient pressure \cite{La3Ni2O7film,(LaPr)3Ni2O7film} has spurred proposals of several effective interacting models for investigating various physical properties \cite{Yue_2025,qiu2025pairingsymmetrysuperconductivityla3ni2o7,shao2025bandstructurepairingnature,cao2025strainengineeredelectronicstructuresuperconductivity}, building on insights from \textit{ab initio} calculations \cite{bhatt2025resolvingstructuraloriginssuperconductivity,Liu_2025,hu2025electronic,Yue_2025,shi2025effectcarrierdopingthickness}. These effective models primarily focus on parameterizing the magnetic correlations based on direct exchange couplings between nickel's $d$-orbitals, while the contribution of oxygen degrees of freedom is neglected. 
In this work, the 11-band $d-p$ Hubbard model, which incorporates both Ni-$3d$ and O-$2p$ orbitals, provides a more comprehensive picture for understanding the relative strengths of superexchange couplings and  distributions of charge carriers in La$_3$Ni$_2$O$_7$ thin films. Crucially, we resolve properties of in-plane and out-of-plane bands exhibiting remarkable different hole concentrations ($n_h$) and correlation strengths. We find that   IR and IT magnetic coupling strengths are more or less comparable, in sharp contrast to the IT coupling  - dominated bulk materials. Furthermore, we observe a pronounced asymmetry in charge distribution upon hole/electron doping. 
In summary, our unified analysis on both magnetic correlations and charge-transfer properties consistently identifies  the interlayer $Ni-3d_{3z^2-r^2}$ orbital couplings as the dominant factor determining the nature of pairing in the bilayer nickelate films. The contribution of the more itinerant Ni-$d_{x^2-y^2}$ orbitals in  superconducting or normal state, however, appear more subtle and complex, warranting further dedicated investigations.

\textit{Acknowledgements.-}
We thank the useful discussion with Zhihui Luo and Wenyuan Qiu. This project was supported by NSFC-12494591, NSFC-12494594, NSFC-92165204, NSFC-92565303, NKRDPC-2022YFA1402802, Guangdong Provincial Key Laboratory of Magnetoelectric Physics and Devices (2022B1212010008), Research Center for Magnetoelectric Physics of Guangdong Province (2024B0303390001), and Guangdong Provincial Quantum Science Strategic Initiative (GDZX2401010).



\bibliographystyle{cplliu}
\bibliography{main}



\end{document}